\documentstyle[12pt]{article}
\newlength{\extraspace}
\newlength{\extraspaces}
\newcommand{\be}{\begin{equation}
\addtolength{\abovedisplayskip}{\extraspaces}
\addtolength{\belowdisplayskip}{\extraspaces}
\addtolength{\abovedisplayshortskip}{\extraspace}
\addtolength{\belowdisplayshortskip}{\extraspace}}
\newcommand{\ee}{\end{equation}}
\newcommand{\ba}{\begin{eqnarray}
\addtolength{\abovedisplayskip}{\extraspaces}
\addtolength{\belowdisplayskip}{\extraspaces}
\addtolength{\abovedisplayshortskip}{\extraspace}
\addtolength{\belowdisplayshortskip}{\extraspace}}
\newcommand{\ea}{\end{eqnarray}}
\newcommand{\nonu}{\nonumber \\[.5mm]}
\newcommand{\tr}{\, {\rm tr} \,}
\newcommand{\e}{\, {\rm e}}
\newcommand{\bra}[1]{\left\langle {#1} \right\vert}
\newcommand{\ket}[1]{\left\vert {#1} \right\rangle}
\newcommand{\VEV}[1]{\left\langle {#1} \right\rangle}
\begin{document}
\addtolength{\baselineskip}{.7mm}
\thispagestyle{empty}
\begin{flushright}
{\sc STUPP}--92--127 \\ February 1992
\end{flushright}
\vspace{6mm}
\begin{center}
{\large{\bf{Schwinger-Dyson Equations of Matrix}\\[3mm]
{Models for Open and Closed Strings}}}\\[15mm]
{\sc Yukihisa Itoh} \\[8mm] and \\[8mm]
{\sc Yoshiaki Tanii}\\[15mm]
{\it Physics Department, Saitama University \\[2mm]
Urawa, Saitama 338, Japan} \\[20mm]
{\bf Abstract}\\[10mm]
\end{center}
{\parbox{130mm}{
We study the Schwinger-Dyson equations of a matrix model for an
open-closed string theory. The free energy with source terms for
scaling operators satisfies the same Virasoro conditions as those of
the pure closed string and is obtained from that of the pure closed
string by giving appropriate nonvanishing background values to all
of the sources.}}
\vfill
\newpage
\setcounter{equation}{0}
Recent studies on the double scaling limit of matrix models
\cite{BRKA} - \cite{DOUGLAS} have provided a nonperturbative
formulation of random surfaces. Such a formulation is necessary
to fully understand two-dimensional quantum gravity and string
theories. Most of these works concern random surfaces without
boundary, which are relevant to pure closed string theories.
In ref.\ \cite{KAZAKOV} a matrix model for random surfaces
with boundaries was introduced and was examined in the Veneziano
limit, where surfaces with handles are ignored. This model can be
regarded as either a zero-dimensional string theory containing both
of open and closed strings or two-dimensional quantum gravity
without matter on surfaces with boundaries.
Subsequently, the double scaling limit of the model was
studied in ref.\ \cite{KOSTOV} and the string equation which
includes effects of surfaces with handles as well as boundaries was
obtained. Matrix models for other open-closed string theories were
discussed in ref.\ \cite{YANG}.
\par
The purpose of this paper is to study further properties of matrix
models for open-closed string theories and compare them with the
pure closed string case. In particular, we use the Schwinger-Dyson
(S-D) loop equations \cite{WADIA} of the matrix model to
derive conditions satisfied by the free energy with source terms
for the scaling operators in the continuum theory. It is a
generalization of the studies in
refs.\ \cite{FUKANA}, \cite{DIVEVE} for the pure closed string to
the open-closed string case. We find that the free energy satisfies
the same Virasoro conditions as those of the pure closed string case
\cite{FUKANA}, \cite{DIVEVE}. The free energy of the open-closed
string theory is obtained from that of the pure closed string theory
by giving appropriate nonvanishing background values to all of the
sources. Therefore the open-closed case and the pure closed case
can be treated in a unified way. As a preparation for these
analysis we also compute one- and two-point correlation functions
explicitly on the sphere and on the disk.
\par
%
%
In order to explain our notations let us briefly review the results
in refs.\ \cite{KAZAKOV}, \cite{KOSTOV}.
The matrix model introduced in ref.\ \cite{KAZAKOV} to
describe both of open and closed strings is
\ba
\e^F &\!\!\! = &\!\!\! \int d \Phi \exp
\left( -N \tr U(\Phi) \right), \nonu
U(\Phi) &\!\!\! = &\!\!\! {1 \over 2} \Phi^2
- {1 \over 4} \lambda \Phi^4
+ \gamma \ln (1-\mu^2\Phi^2),
\label{model}
\ea
where $\Phi$ is an $N \times N$ hermitian matrix and
$\lambda$, $\gamma$ and $\mu$ are real parameters. Due to the U($N$)
invariance the integral (\ref{model}) is reduced to the integral
over the eigenvalues $\phi_1, \cdots, \phi_N$ of the matrix. The
eigenvalue integral can be evaluated by the method of orthogonal
polynomials. The polynomial $P_n (\phi)$ of order $n$
$(= 0, 1, 2, \cdots)$ is defined by
\be
\VEV{m | n}
\equiv \int d \phi \exp \left( -N \tr U(\phi) \right)
P_m (\phi) P_n (\phi) = \delta_{m, n}.
\label{pol}
\ee
The free energy $F$ can be expressed in terms of the matrix elements
$r_n$ of $\phi$
\be
\bra{m} \hat \phi \ket{n} = \sqrt{r_m} \, \delta_{m, n+1}
+ \sqrt{r_n} \, \delta_{n, m+1}.
\label{phivev}
\ee
{}From the identity
$(2n+1) N^{-1} = \bra{n} \hat \phi U'(\hat \phi) \ket{n}$
one obtains a recursion relation for $r_n$
\ba
(2n+1) N^{-1}
&\!\!\! = &\!\!\! \; r_n \left[1-\lambda (r_{n-1} + r_n + r_{n+1})
\right] \nonu
&\!\!\! &\!\!\! + r_{n+1} \left[1-\lambda (r_n + r_{n+1} + r_{n+2})
\right]
+ 2 \gamma - 2 \gamma \bra{n} {1 \over 1 - \mu^2 \hat \phi^2}
\ket{n}.
\label{rrelation}
\ea
\par
The double scaling limit is achieved by introducing a lattice spacing
$a$ and by scaling the parameters as
\be
N = a^{-{5 \over 2}}, \quad \gamma = a^{5 \over 2} \Gamma, \quad
\lambda = {1 \over 12} (1-t a^2), \quad
\mu = {1 \over 2\sqrt 2}\e^{-aM},
\label{scalingp}
\ee
where $\Gamma$ is the renormalized open string coupling constant.
We have put the closed string coupling constant, which could appear
in the first equation of eq.\ (\ref{scalingp}), to be unity.
The parameters $t$ and $M$ are the renormalized bulk and boundary
cosmological constants and are conjugate to the area and the
boundary length of the surfaces respectively. In the limit
$a \rightarrow 0$ the recursion relation (\ref{rrelation}) is
reduced to \cite{KOSTOV}
\be
\xi = u^2 - {1 \over 3} u'' + 2 \Gamma R(\xi, \xi; 2M),
\label{stringeq}
\ee
where we have defined
\be
x = {n \over N}, \quad 1-12\lambda x = a^2 \xi, \quad
r_n = r(x) = {1 \over 6\lambda} \left( 1 -a u(\xi) \right).
\label{scalingv}
\ee
The diagonal element of the Gel'fand-Dikii resolvent \cite{GEDI}
\be
R(\xi, \xi; 2M) = \bra{\xi}
{1 \over - \partial_\xi^2 + u(\xi) + 2M} \ket{\xi}
\label{resolvent}
\ee
satisfies
\be
1 = - 2 R R'' + R'^2 + 4 (u + 2M) R^2
\label{resolventeq}
\ee
and has an expansion
\be
R(\xi, \xi; 2M)
= \sum_{k=0}^\infty {R_k[u] \over (2M)^{k+{1 \over 2}}},
\label{resexp}
\ee
where the coefficients $R_k[u]$ are functions of $u$ and its
derivatives:
\be
R_0 = {1 \over 2}, \quad
R_1 = - {1 \over 4} u, \quad
R_2 = - {1 \over 16} u'' + {3 \over 16} u^2, \quad \cdots.
\label{coeff}
\ee
Eq.\ {\ref{stringeq}) is called the string equation.
The function $u(t, M)$ is shown to be the second derivative of
the free energy $F(t, M)$ with respect to $t$.
Substituting eq.\ (\ref{stringeq}) into eq.\ (\ref{resolventeq})
one obtains the differential equation of fourth order for $u$.
\par
The Veneziano limit, in which surfaces with an arbitrary number of
boundaries but no handle are allowed, is obtained by dropping the
derivatives of $u$ in eq.\ (\ref{stringeq}). The string equation
reduces to
\be
\xi = u^2 + {\Gamma \over \sqrt{2M+u}}.
\label{veneziano}
\ee
It can be solved as a power series in $\Gamma$
\be
u(\xi) = \sqrt \xi - {\Gamma \over 2 \sqrt \xi
\sqrt{2M + \sqrt \xi}} + O(\Gamma^2).
\label{spheredisk}
\ee
The power of $\Gamma$ in each term of the solution represents
the number of boundaries on the corresponding surface.
Therefore, the first and the second terms in eq.\ (\ref{spheredisk})
are contributions from the sphere and the disk respectively.
\par
%
%
Now we shall consider correlation functions of operators.
In string theories containing open as well as closed strings there
exist two kinds of operators: the closed string emission vertex
operators and the open string emission vertex operators.
We will consider only the closed string operators.
In the matrix model they are linear combinations of
\be
W_n = {1 \over N} \tr \Phi^n \quad (n = 0, 1, 2, \cdots)
\label{operator}
\ee
and scattering amplitudes of closed strings are given by
their correlation functions. It is convenient to introduce the
generating functions of connected Green's functions
\ba
G_K (z_1, z_2, \cdots, z_K)
&\!\!\! = &\!\!\! \sum_{n_i = 0}^\infty z_1^{-n_1-1}
\cdots z_K^{-n_K-1} \VEV{W_{n_1} \cdots W_{n_K}}_c \nonu
&\!\!\! = &\!\!\! \VEV{{1 \over N} \tr {1 \over z_1 - \Phi} \,
\cdots \, {1 \over N} \tr {1 \over z_K - \Phi}}_c.
\label{genfunc}
\ea
They are the Laplace transforms of correlation functions of
macroscopic loops with fixed lengths.
\par
In the double scaling limit $a \rightarrow 0$ with
eqs.\ (\ref{scalingp}), (\ref{scalingv}) and
\be
z_i = 2 \sqrt 2 \e^{a \zeta_i},
\label{zscaling}
\ee
they behave as
\ba
G_1 (z) &\!\!\! = &\!\!\! G_1^{\rm non} (z) + a^{3 \over 2}
g^{(1)}(\zeta), \nonu
G_2 (z_1, z_2) &\!\!\!
= &\!\!\! G_2^{\rm non} (z_1, z_2)
+ a^3 g^{(2)}(\zeta_1, \zeta_2), \nonu
G_K (z_1, \cdots, z_K) &\!\!\!
= &\!\!\! a^{{3 \over 2}K} g^{(K)}(\zeta_1, \cdots, \zeta_K).
\label{gscaling}
\ea
The functions $G_1^{\rm non}$ and $G_2^{\rm non}$ are nonuniversal
parts of the scaling limit and are not relevant in the continuum
theory. The nonuniversal terms exist only for one-point function on
the sphere and on the disk, and two-point function on the sphere.
This can be understood as follows. The correlation functions of
macroscopic loops can be regarded as partition functions on surfaces
with holes. The partition function for a surface with the Euler
number $\chi$ is expressed as an integral over area of the surface
\cite{DDK}:
\be
Z_\chi (t, M) \sim \int_0^\infty dA A^{-{5 \over 4}\chi-1}
\left[ 1 + \sum_{n=1}^\infty c_n
\left( M A^{1 \over 2} \right)^n \right] \e^{- t A}
\label{areaint}
\ee
for the present case of pure gravity, where $c_n$ are constants
independent of $t, \; M$ and $A$. The integral is convergent
for surfaces with $\chi < 0$ and is proportional to
$t^{{5 \over 4}\chi}$ times a power series in $M \, t^{-1/2}$.
This is the $t$ dependence of the universal terms in
eq.\ (\ref{gscaling}). On the other hand, for surfaces
with $\chi \geq 0$, i.e. the sphere with one or two holes and the
disk with one hole, the integral is divergent at $A \sim 0$ and has
nonuniversal terms of polynomial in $t$. The universal parts in
eq.\ (\ref{gscaling}) can be expanded as \cite{GRMI}
\be
g^{(K)}(\zeta_1, \cdots, \zeta_K)
= \sum_{n_i=0}^\infty \zeta_1^{-n_1-{3 \over 2}} \cdots
\zeta_K^{-n_K-{3 \over 2}} g_{n_1, \cdots, n_K}.
\label{univexp}
\ee
The coefficients $g_{n_1, \cdots, n_K}$ of the expansion can be
regarded as correlation functions of scaling
operators \cite{GRMI} $O_{n_1}, \cdots, O_{n_K}$ of the
continuum theory.
\par
As explicit examples we shall evaluate the one- and two-point
functions on the sphere and on the disk. In the scaling limit they
are given by \cite{GRMI}
\ba
G_1 (z) &\!\!\! = &\!\!\! {1 \over N} \int_0^1 dx
\bra{x} {1 \over z - \hat\phi} \ket{x}, \nonu
G_2 (z_1, z_2) &\!\!\! = &\!\!\! {1 \over N^2} \int_0^1 dx_1
\int_1^\infty dx_2
\bra{x_2} {1 \over z_1 - \hat\phi} \ket{x_1}
\bra{x_1} {1 \over z_2 - \hat\phi} \ket{x_2}.
\label{onetwo}
\ea
The matrix elements in eq.\ (\ref{onetwo}) can be evaluated by
using the fact that in the scaling limit $\hat\phi$ is represented
by a second order differential operator
\be
\hat\phi = 2\sqrt{2} - \sqrt{2} a \left(
- \partial_\xi^2 + u(\xi) \right) + O(a^{3 \over 2}).
\label{diffop}
\ee
To obtain the correlation functions on the sphere and on the disk
we substitute the solution (\ref{spheredisk}) for $u$ into
eq.\ (\ref{diffop}) and neglect derivatives of $u$ in
eq.\ (\ref{onetwo}). After some calculation we find
that the one- and two-point functions on the sphere and on the disk
have the forms in eq.\ (\ref{gscaling}). The universal terms are
given by
\ba
g^{(1)}(\zeta) &\!\!\!
= &\!\!\! \; -{\sqrt{2} \over 3}
\left( 2\zeta + \sqrt{t} \right)^{3 \over 2}
+ 2\sqrt{2} \, \zeta \left( 2\zeta + \sqrt{t} \right)^{1 \over 2}
-{8 \over 3} \, \zeta^{3 \over 2}
+ {1 \over 4} \, t \, \zeta^{-{1 \over 2}} \nonu
&\!\!\! &\!\!\! + {\Gamma \over 4\sqrt{2}} \, {1 \over \zeta - M}
\left( \sqrt{M \over \zeta}
- \sqrt{2M+\sqrt{t} \over 2\zeta+\sqrt{t}} \right), \nonu
g^{(2)}(\zeta_1, \zeta_2)
&\!\!\! = &\!\!\! \; {1 \over 8}{1 \over \sqrt{
(2\zeta_1+\sqrt{t})(2\zeta_2+\sqrt{t})} \left(
\sqrt{2\zeta_1+\sqrt{t}}+\sqrt{2\zeta_2+\sqrt{t}}\right)^2} \nonu
&\!\!\! &\!\!\! - {1 \over 32}{1 \over \sqrt{\zeta_1 \zeta_2}
(\sqrt{\zeta_1}+\sqrt{\zeta_2})^2} \nonu
&\!\!\! &\!\!\! + {\Gamma \over 32 \sqrt{t}} \,
(2\zeta_1+\sqrt{t})^{-{3 \over 2}}
(2\zeta_2+\sqrt{t})^{-{3 \over 2}} (2M+\sqrt{t})^{-{1 \over 2}}.
\label{onetwouniv}
\ea
It is easy to see that they can be expanded as in
eq.\ (\ref{univexp}). The nonuniversal terms, which we will need
later, are
\ba
G_1^{\rm non} (z) &\!\!\! = &\!\!\! {\sqrt 2 \over 3}
- \sqrt 2 \, a \, \zeta
+ a^{3 \over 2} \left ( {8 \over 3} \, \zeta^{3 \over 2}
- {1 \over 4} \, t \, \zeta^{-{1 \over 2}} \right )
+ {\Gamma \over 4 \sqrt 2} \, a^{3 \over 2}
{1 \over \sqrt \zeta (\sqrt \zeta + \sqrt M)}, \nonu
G_2^{\rm non} (z_1, z_2) &\!\!\! = &\!\!\! {1 \over 32} a^3
{1 \over \sqrt{\zeta_1 \zeta_2} (\sqrt \zeta_1 + \sqrt \zeta_2 )^2}.
\label{nonuniversal}
\ea
In eqs.\ (\ref{onetwouniv}) and (\ref{nonuniversal}) the terms
independent of $\Gamma$ and those
linear in $\Gamma$ are contributions from the sphere and the disk
respectively. The sphere contributions in the nonuniversal
parts (\ref{nonuniversal}) were given in ref.\ \cite{FUKANA}.
\par
%
%
We now proceed to study the S-D equations and derive constraints
satisfied by the free energy with source terms for the scaling
operators. We closely follow the approach in
ref.\ \cite{FUKANA}. By a change of integration variables
in the matrix integral we obtain the S-D
equations \cite{WADIA} for the connected
Green's functions of the operators (\ref{operator})
\ba
&\!\!\! &\!\!\! \VEV{W_{m+1} \prod_{j=1}^K W_{n_j}}_c
- \lambda \VEV{W_{m+3} \prod_{j=1}^K W_{n_j}}_c
- 2 \gamma \sum_{n=1}^\infty \mu^{2n}
\VEV{W_{m+2n-1} \prod_{j=1}^K W_{n_j}}_c \nonu
&\!\!\! &\!\!\! = {1 \over N^2} \sum_{j=1}^K n_j
\VEV{\prod_{k=1}^{j-1} W_{n_k}
W_{n_j+m-1} \prod_{l=j+1}^K W_{n_l}}_c
+ \sum_{j=0}^{m-1} \VEV{W_j W_{m-j-1} \prod_{k=1}^K W_{n_k}}_c \nonu
&\!\!\! &\!\!\! \quad + \sum_{j=0}^{m-1} \sum_{S \subseteq \{1, 2,
\cdots, K\}}
\VEV{W_j \prod_{k\in S} W_{n_k}}_c
\VEV{W_{m-j-1} \prod_{l\in \bar S} W_{n_l}}_c.
\label{sdeqw}
\ea
In terms of the generating functions (\ref{genfunc}) the S-D
equations (\ref{sdeqw}) can be rewritten as
\ba
&\!\!\! &\!\!\! \left( z - \lambda z^3 - {2 \gamma \mu^2 z \over 1
- \mu^2 z^2} \right)
G_{K+1} (z, z_1, \cdots, z_K)
+ (\lambda z^2 -1) G_K (z_1, \cdots, z_K) \nonu
&\!\!\! &\!\!\! \quad + \lambda \sum_{n_i=0}^\infty z_1^{-n_1-1}
\cdots z_K^{-n_K-1}
\left[ z \VEV{W_1 \prod_{j=1}^K W_{n_j}}_c
+ \VEV{W_2 \prod_{j=1}^K W_{n_j}}_c \right] \nonu
&\!\!\! &\!\!\! \quad + {2 \gamma \mu^2 \over 1 - \mu^2 z^2}
\sum_{m, n_i =0}^\infty \mu^{2m} z_1^{-n_1-1} \cdots z_K^{-n_K-1}
\VEV{W_{2m} \prod_{j=1}^K W_{n_j}}_c \nonu
&\!\!\! &\!\!\! \quad + {2 \gamma \mu^3 z \over 1 - \mu^2 z^2}
\sum_{m, n_i =0}^\infty \mu^{2m+1} z_1^{-n_1-1} \cdots z_K^{-n_K-1}
\VEV{W_{2m+1} \prod_{j=1}^K W_{n_j}}_c \nonu
&\!\!\! &\!\!\! = G_{K+2} (z, z, z_1, \cdots, z_K) \nonu
&\!\!\! &\!\!\! \quad + \sum_{n=0}^K
\sum_{{{S_1 = \{i_1, \cdots, i_n\} \atop
S_2 = \{i_{n+1}, \cdots, i_K\}} \atop
S_1 \cap S_2 =\phi } \atop
S_1 \cup S_2 = \{1, 2, \cdots, K\}}
G_{n+1} (z, z_{i_1}, \cdots, z_{i_n})
G_{K-n+1} (z, z_{i_{n+1}}, \cdots, z_{i_K}) \nonu
&\!\!\! &\!\!\! \quad + {1 \over N^2} \sum_{j=1}^K {\partial \over
\partial z_j}
{G_K (z_1, \cdots, z_K)
- G_K (z, z_1, \cdots, \hat z_j, \cdots, z_K) \over z_j - z},
\label{sdeqg}
\ea
where $\hat z_j$ in the last term means an omission of $z_j$ in
the arguments. In the double scaling limit we substitute
eqs.\ (\ref{gscaling}) and (\ref{nonuniversal}) into
eq.\ (\ref{sdeqg}) and obtain
equations for the functions $g^{(K)}$ in eq.\ (\ref{gscaling}).
Expanding such equations in $\zeta, \zeta_1, \cdots, \zeta_K$ and
looking at terms of negative powers in $\zeta$, we obtain relations
among the expansion coefficients $g_{n_1, \cdots, n_K}$
in eq.\ (\ref{univexp}).
\par
To write down such relations in a compact form we introduce
\ba
{1 \over 2} \, g(\mu_0, \mu_1, \cdots)
&\!\!\! = &\!\!\! \sum_{n_i=0}^\infty {\mu_0^{n_0} \over n_0!}
{\mu_1^{n_1} \over n_1!} \cdots
g_{\, \underbrace{\scriptstyle{0, \, \cdots, \, 0 \, }}_{n_0} \, ,
\, \underbrace{\scriptstyle{1, \, \cdots, \, 1 \, }}_{n_1}
\, , \cdots}, \nonu
\tau (\mu_0, \mu_1, \cdots) &\!\!\!
= &\!\!\! \exp \left( {1 \over 2} g(\mu_0, \mu_1, \cdots) \right).
\label{free}
\ea
As was explained in ref.\ \cite{FUKANA}
$g(\mu_0, \mu_1, \cdots)$ can be regarded as the free energy with
source terms $\mu_0 O_0 + \mu_1 O_1 + \cdots$ for the scaling
operators. A factor ${1 \over 2}$ in front of
$g(\mu_0, \mu_1, \cdots)$ in eq.\ (\ref{free}) is due to the fact
that one should consider only the even operators $W_{2n}$ when
the potential $U(\Phi)$ is an even function. The relations among
$g_{n_1, \cdots, n_K}$ are summarized as differential equations
on $\tau$
\ba
128 {\partial \tau \over \partial \mu_1}
&\!\!\! + &\!\!\! 6\sqrt{2} \, \Gamma \sum_{k=1}^\infty
M^{-k-{1 \over 2}}
{\partial \tau \over \partial \mu_{k-1}}
+ {3 \over 2} \sum_{k=1}^\infty (2k+1) \mu_k
{\partial \tau \over \partial \mu_{k-1}} \nonu
&\!\!\! &\!\!\! \qquad\qquad\qquad\qquad\qquad\qquad
+ {3 \over 128} \left(
\mu_0 - 8t + {4\sqrt{2} \, \Gamma \over \sqrt{M}}
\right)^2 \tau = 0, \nonu
128 {\partial \tau \over \partial \mu_2}
&\!\!\! - &\!\!\! 12 t {\partial \tau \over \partial \mu_0}
+ 6 \sqrt{2} \, \Gamma \sum_{k=0}^\infty M^{-k-{1 \over 2}}
{\partial \tau \over \partial \mu_k}
+ {3 \over 2} \sum_{k=0}^\infty (2k+1) \mu_k
{\partial \tau \over \partial \mu_k}
+ {3 \over 16} \tau = 0, \nonu
128 {\partial \tau \over \partial \mu_{p+3}}
&\!\!\! - &\!\!\! 12 t {\partial \tau \over \partial \mu_{p+1}}
+ 6 \sqrt{2} \, \Gamma \sum_{k=0}^\infty M^{-k-{1 \over 2}}
{\partial \tau \over \partial \mu_{k+p+1}} \nonu
&\!\!\! + &\!\!\! {3 \over 2} \sum_{k=0}^\infty (2k+1) \mu_k
{\partial \tau \over \partial \mu_{k+p+1}} + 24 \sum_{r=0}^p
{\partial^2 \tau \over \partial \mu_r \partial \mu_{p-r}} = 0
\quad (p \ge 0).
\label{diffeq}
\ea
The terms depending on $\Gamma$ are effects of surfaces with
boundaries. If we set $\Gamma = 0$, we obtain the equations for the
pure closed string derived in refs.\ \cite{FUKANA}, \cite{DIVEVE}.
Actually, the terms depending on $\Gamma$ can be absorbed by a
shift of the sources
\be
\mu'_k = \mu_k + {4 \sqrt 2 \over 2k+1} \Gamma M^{-k-{1 \over 2}}.
\label{shift}
\ee
Therefore, in terms of the shifted sources $\mu'_k$ the equations
have the same form as those of the pure closed string case.
It is interesting to note that string theories containing both of
open and closed strings can be described by the same equations as
those of pure closed string theories. The difference of two cases
arises when one chooses background values of the sources, to which
small perturbations are added. Just as one can interpolate between
different multicritical points in pure closed string theories by
switching on some of the sources, one can also interpolate between
pure closed string theories and open-closed string theories by
giving appropriate nonvanishing background values to all of the
sources.
\par
{}From the fact that the scaling operator $O_0$ corresponds to the
cosmological term, we require as in the pure closed string
case \cite{FUKANA}
\be
{\partial \tau \over \partial \mu_0}
= - {1 \over 8} {\partial \tau \over \partial t},
\label{cosmological}
\ee
which is consistent with eq.\ (\ref{diffeq}).
Eq.\ (\ref{cosmological}) shows
that $\tau$ depends on $\mu_0$ and $t$ only through a combination
$\mu_0 - 8t$.
\par
By changing the variables to
\be
-2^{k+3} x'_{2k+1}
= \mu'_k - 8t \delta_{k, 0} + {256 \over 15} \delta_{k, 2},
\label{xmu}
\ee
eq.\ (\ref{diffeq}) is rewritten as the Virasoro conditions
on $\tau$ \cite{FUKANA}, \cite{DIVEVE}
\ba
L_n \tau &\!\!\! = &\!\!\! 0 \qquad (n=-1, 0, 1, \cdots), \nonu
2 L_n &\!\!\! = &\!\!\! {1 \over 2} \sum_{p+q=-2n} p q x'_p x'_q
+ \sum_{p-q=-2n} p x'_p \partial'_q
+ {1 \over 2} \sum_{p+q=2n} \partial'_p \partial'_q
+ {1 \over 8} \delta_{n, 0},
\label{virasoro}
\ea
where $p, \; q$ are positive odd integers.
By assuming that $\tau$ is the $\tau$-function of the KdV hierarchy
\cite{DATE}, the $x'_1$-derivative of the condition
$L_{-1} \tau = 0$ gives the string equation
\cite{FUKANA}, \cite{DIVEVE}
\be
{1 \over 2} x'_1 + \sum_{k=1}^\infty (2k+1) x'_{2k+1} R_k [u] = 0.
\label{multistringeq}
\ee
In terms of the variables $x_k$ related to $\mu_k$ by the same
relation as eq.\ (\ref{xmu}), the string
equation (\ref{multistringeq}) becomes
\be
{1 \over 2} x_1 + \sum_{k=1}^\infty (2k+1) x_{2k+1} R_k [u]
- \Gamma R(x_1, x_1; 2M) = 0,
\label{multistringeqx}
\ee
where we have used the formulae (\ref{resexp}) and (\ref{coeff}).
This is a generalization of the string equation (\ref{stringeq})
for the pure gravity to more general theories.
The $k$-th multicritical theory is obtained by setting all $x$'s
to zero except $x_1$ and $x_{2k+1}$.
\par
}
\end{document}